\documentstyle[aps,twocolumn]{revtex}
\begin{document}

\input epsf

\twocolumn[\hsize\textwidth\columnwidth\hsize\csname      
@twocolumnfalse\endcsname                                 

\title{The calculation of the impurity entropy for the exactly 
solvable two-channel Kondo model.}

\author{A.V. Rozhkov}

\address{
Department~of~Physics, University~of~California,~San~Diego, 9500~Gilman~Drive,
La Jolla, CA~92093-0354, USA
}

\maketitle

\begin{abstract}

In this paper the entropy of the two-channel Kondo impurity is calculated.
We demonstrate that the exactly solvable system of finite size can be
described in terms of free fermions. None of these fermions is Majorana
fermion. This allows us to find the impurity entropy $S(T, L)$ for
arbitrary values of $T$ and $L$. As found before, $S(T=0, L=\infty)$ equals
to $(1/2)\ln 2$. For finite $L$ the entropy shows a cross-over to the
infinite-size behavior at $T>T_L=v_F/L$, in agreement with results of
other researchers.

\end{abstract}


\vskip2pc] \narrowtext    

\draft

\section{Introduction}
Recent investigations of macroscopic quantum phenomenon, 1D quantum wires
and Kondo effect discover that the zero-temperature entropy of some systems
possesses an anomaly \cite{paco,lesage,emery,tsvelik}.
Specifically, zero-temperature entropy 
of the two-channel Kondo impurity was found to be: 
\begin{equation}
S(T=0)={1\over 2}\ln 2. \label{entr}
\end{equation}

This result is in obvious contradiction with the fundamental fact:
for any system which exists in nature it is assumed that
$\exp(S(T=0))$ is the degeneracy of the ground state, i.e.
integer number. Equation (\ref{entr}), however, implies that the degeneracy
equals to $\sqrt{2}$. 

It was pointed out later in the context of Kondo model 
\cite{affleck,coleman} that
the entropy value $(1/2)\ln 2$ is obtained in the thermodynamics limit where
system size $L$ is sent to infinity. For a finite size system a
non-zero temperature scale can be defined in the following way:
${T_L}=v_F/L$. Thermodynamic limit calculations are not valid below $T_L$.
Mathematically, this situation can be described as follows \cite{affleck}:
\begin{equation}
\lim_{L\rightarrow \infty} \lim_{T\rightarrow 0} S = \ln 2 \ne
\lim_{T\rightarrow 0} \lim_{L\rightarrow \infty} S = {1\over 2} \ln 2.
\label{lim}
\end{equation}
Only the first of these two limits is physically relevant \cite{expl} and it
gives reasonable result.

As shown in \cite{emery} (further cited as I) two-channel Kondo model is
exactly solvable at particular values of coupling constants. In this paper
we study that exactly solvable model in the limit of finite $L$.
A description of the finite size exactly solvable system in terms of 
free Dirac fermions is found. Majorana fermion which appears in $L=\infty$
system combines with another Majorana fermion and forms Dirac fermion. In
$L=\infty$ limit that second Majorana fermion remains hidden in the
continuous spectrum. Numerical calculation of the spectrum of
the free fermions allows us to obtain the impurity entropy $S$ for
arbitrary values of $T$ and $L$. As a function of $T$ and $L$ the entropy
demonstrates a cross-over at $T\sim T_L$, in agreement with the theory of
critical phenomena. For $T>T_L$ finite size effects are unimportant while
below $T_L$ the impurity decouples from the Fermi
sea. We confirm this by both numerical and analytical calculations.

The paper is organized as follows. In Section II we calculate the spectrum
of the finite size version of Kondo model. As we said this model can be
reduced to a set of free fermions. In Section III the
thermodynamics of the model is discussed. Section IV contains the
conclusions.

\section{Spectrum of two-channel Kondo impurity}

In this part we are going to diagonalize the hamiltonian of two-channel
Kondo impurity placed in the box of size $L$.

In paper I two-channel Kondo hamiltonian
\begin{equation}
H=i{v_F}\sum_{i, \alpha=1}^2\int_{-L}^{L}dx\psi_{i\alpha}^{\dagger}(x)
\partial_{x}\psi_{i\alpha}(x) + H_I, \label{ham}
\end{equation}
\begin{equation}
H_{I}={1\over 2}\sum_{i,\alpha,\beta=1}^{2}\sum_{\lambda=x,y,z}
J_{\lambda}\tau_{\lambda}\sigma_{\alpha\beta}^{\lambda}
\psi_{i\alpha}^\dagger(0)\psi_{i\beta}(0) \label{HI}
\end{equation}
was studied in the limit of infinite $L$.
Here $\psi$ -- left-moving electrons, $\alpha, \beta$ are electron spin
indices, index $i$ is electron "flavor", $\tau^\lambda$ is the impurity
pseudospin. $H_I$ describes the coupling of the electron spin to the
impurity pseudospin. Coupling constants $J_{x,y,z}$ are chosen in such a
way that $J_x=J_y, J_z=\pi v_F>0$. 

Following the procedure described in I
(bosonization and re-fermionization) this hamiltonian can be transformed to:
\begin{equation}
H_{new}=i{v_F}\int_{-L}^{L} dx \Psi^\dagger(x)\partial_{x}\Psi(x)+
\label{Hnew}
\end{equation}
$$
{J_x\over{\sqrt{2\pi a}}}\left[
\Psi^\dagger(0)+\Psi(0)\right](d^\dagger-d), 
$$
where $\Psi$ is some new fermionic field, $d$ is a fermionic operator
representing the impurity, $a$ is a smallest length in the system (unit cell
size).
Expressing this hamiltonian in terms of $c_k$'s, Fourier components of
$\Psi$, we obtain:
\begin{equation}
H_{new}=\sum_{k} {v_F}kc_{k}^{\dagger}c_k +
\Delta\sum_{k}(c_{k}^{\dagger}+c_{k})(d^{\dagger}-d), \label{H}
\end{equation}
\begin{equation}
\Delta={J_x\over\sqrt{4\pi a L}}.	\label{Del}
\end{equation}
Hamiltonian (\ref{Hnew}) written in terms of Fourier components (\ref{H})
will be a starting point for our calculations.

To proceed further let us introduce two operators:
\begin{eqnarray}
D_{1}={1\over \sqrt{2}}\left(d^\dagger+d\right),\\
D_{2}={i\over \sqrt{2}}(d^\dagger-d).
\end{eqnarray}
Obviously, $D_1$ and $D_2$ satisfy the following relations:
\begin{eqnarray}
D_{1,2}^\dagger=D_{1,2}, \label{conj}\\
\{D_1;D_1\}=\{D_2;D_2\}=1, \label{idemp}\\
\{D_{1};D_{2}\}=0, \label{ferm}
\end{eqnarray}
Though (\ref{ferm}) is usual property of fermionic operators the equalities
(\ref{conj}) and (\ref{idemp}) define Majorana fermion operators.
It was pointed out in I that $D_1$ commutes with the hamiltonian:
\begin{equation}
\left[H_{new}; D_1\right]=0.
\end{equation}
The anomalous entropy behavior in case of infinite $L$ was attributed to
the presence of this
Majorana fermion commuting with (\ref{H}). We will see below that there is
no Majorana fermion and no anomalous entropy for the finite size
system.

Now let us diagonalize the hamiltonian (\ref{H}). Since it is quadratic in
fermionic operators, we will look for a set of eigenoperators of the
form:
\begin{equation}
f^\dagger=i\sum_{k}\left( \tilde{\gamma}_{k}c_k^\dagger  
+ \gamma_k c_k \right) + \delta D_2,	\label{deff}
\end{equation}
such that
\begin{equation}
\left[H_{new};f^\dagger\right]=\omega f^\dagger.	\label{Hf}
\end{equation}

Conditions (\ref{deff}-\ref{Hf}) give rise to an eigenvalue problem for
$\gamma$'s and $\delta$:
\begin{eqnarray}
\omega\tilde{\gamma}_k=\epsilon_k\tilde{\gamma}_k-\sqrt{2}\Delta\delta, 
\label{gam}\\
\omega\gamma_k=-\epsilon_k\gamma_k-\sqrt{2}\Delta\delta, \\
\omega\delta=-\sqrt{2}\Delta\sum_{k}\left(\tilde{\gamma}_k+
\gamma_k\right).\label{delta}
\end{eqnarray}
Here $\epsilon_k={v_F}k$.

The solution of this set of equations is:
\begin{eqnarray}
\tilde{\gamma}_k=-\delta{\sqrt{2}\Delta\over\omega-\epsilon_k},\\
\gamma_k=-\delta{\sqrt{2}\Delta\over\omega+\epsilon_k},\\
\omega=2\Delta^2\sum_{k} 
{1\over\omega+\epsilon_k}+{1\over\omega-\epsilon_k}. \label{char}
\end{eqnarray}
The condition $\{f^\dagger;f\}=1$ fixes the value of $\delta$.

The solution shown above allows one to verify the following property of
symmetry: equations (\ref{gam}-\ref{delta}) are invariant 
under the transformation
$\omega\rightarrow-\omega, \tilde{\gamma}_k\rightarrow-\gamma_k,
\gamma_k\rightarrow-\tilde{\gamma}_k, \delta\rightarrow\delta.$ 
Another way of expressing this symmetry is:
\begin{equation}
f^\dagger_\omega=f_{-\omega}.	\label{sym}
\end{equation}
This implies that $\omega>0$ solutions of equations (\ref{gam}-\ref{delta})
constitute fermionic spectrum of the hamiltonian (\ref{H}) while $\omega<0$
solutions are mirror image of that spectrum. In other words, $\omega>0$
eigenvectors give us fermionic creation operators $f^\dagger_\omega$ while
$\omega<0$ eigenvectors give us destruction operators: $f^\dagger_
{-|\omega|}=f_{|\omega|}$.

There is another remarkable property of equations (\ref{gam}-\ref{delta}):
$\omega=0$ solution always exists. This can be readily seen from 
(\ref{char}). The
following is true for $\omega=0$ eigenoperator (see (\ref{sym}) and
(\ref{Hf})):
\begin{eqnarray}
f^\dagger_{\omega=0}=f_{\omega=0}, \label{f0}\\
\left[H_{new}; f_{\omega=0}\right]=0. 	\label{com}
\end{eqnarray}

Equations (\ref{f0}), (\ref{com}) show us that $f_{\omega=0}$ 
is the Majorana operator (see properties (\ref{conj}) and (\ref{idemp}))
commuting with the hamiltonian (\ref{H}). Having two Majorana
operators we can define two other operators:
\begin{eqnarray}
F={1\over\sqrt{2}}\left(if_{\omega=0}+D_1\right), \\
F^\dagger={1\over\sqrt{2}}\left(-if_{\omega=0}+D_1\right).
\end{eqnarray}
These two operators are annihilation and creation operators for some new
fermion:
\begin{eqnarray} 
\{F^\dagger ; F\}=1, \\
\{F; f_{\omega}\}=0, \omega\ne 0.
\end{eqnarray}
Since both $f_{\omega=0}$ and $D_1$ commute with hamiltonian, the
new fermion has zero energy:
\begin{equation}
\left[H_{new}; F\right]=0.
\end{equation}
Thus, the set of eigenoperators of $H_{new}$ consists of $f_{\omega}$ for
$\omega>0$, $F$ and their hermitian conjugates. All of them are
usual fermionic operators. We have a description of our system in terms of
free fermions with no Majorana fermion. 
The presence of the zero-energy fermion $F$ implies that $S(T=0,L)=\ln2$.

\section{Free energy of the Kondo impurity}

In order to calculate the entropy let us consider free energy of our
system. In the previous section we have shown that our system can be
described in terms of free fermions. Therefore, the impurity free energy is:
\begin{equation}
{\cal F}=-T\ln 2 - T\sum_{\omega>0} \ln \left( 1+e^{-{\omega/T}}\right)
-{\cal F}_0. \label{freeE}
\end{equation}
where ${\cal F}_0$ is the free energy without the impurity. 
The impurity entropy is:
\begin{equation}
S=-{\partial {\cal F}\over \partial T}.
\end{equation}
We solve (\ref{char}) numerically and find the entropy for different
$T$. Three curves for three different values of $L$ are plotted
on fig.1: $L_a<L_b<L_c=\infty$. One can see from the graph that the system
has a cross-over at $T_L$: as $T$ goes below $T_L$ the impurity decouples
from the Fermi sea due to the presence of the finite size gaps in the
excitation spectrum of the conducting electrons. It is clearly seen on
fig.1 that the limiting value of $S(T,L)$ at $T=0$ and $L=\infty$ depends
on the path in $(T,L)$ plane, in agreement with (\ref{lim}).

\begin{figure} [!t]
\centering
\leavevmode
\epsfxsize=8cm
\epsfysize=7cm
\epsfbox[08 -14 592 718] {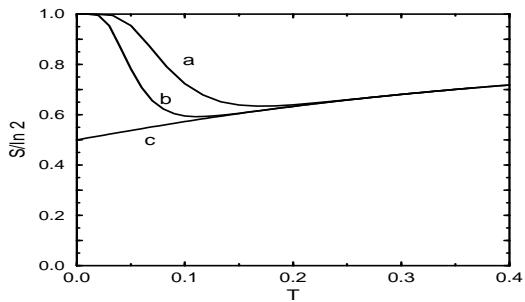}
\caption[]
{
\label{fig1}
The impurity entropy {\it vs.} temperature (in the units of
$\Gamma$) for different values of $L$: $L_a<L_b<L_c=\infty$, $L_a/L_b=0.6$.}
\end{figure}

There is one peculiarity on the graphs of the entropy for the finite
systems.
The entropy of the impurity decreases as a function of temperature below
$T_L$. There is no contradiction here with the fundamental criterion of
thermodynamical stability: the
impurity entropy is not the entropy of any physical system but rather a
difference between the entropy of the system with Kondo impurity and the
entropy of free electrons without impurity. For both of these two systems
the entropy increases with temperature, but their difference does not.

We see form fig.1 that the result for $L=\infty$ works fine as high 
temperature approximation. In order to show this analytically let us
calculate high temperature expression for the impurity free energy
(\ref{freeE}).

In the finite size limit the spectrum consists of a set of
eigenvalues defined by (\ref{char}). The characteristic distance between
two neighboring eigenvalues is $T_L=v_F/L$. However, for $T\gg T_L$ one can
expect that discreteness of the spectrum does not play any role and
it can be approximated by continuum.

In order to study continuous version of our model we are going to obtain
the density of states $\rho(\omega)$.
To do so we will consider the following auxiliary function
\begin{equation}
h(\omega)=\omega\left[\omega-2\Delta^2\sum_{k}{1\over\omega+\epsilon_k} +
{1\over\omega-\epsilon_k}\right]\times
\end{equation}
$$
\prod_{k}(\omega+\epsilon_k)(\omega -
\epsilon_{k}).
$$
As it follows from (\ref{char})
this function has two important properties: if $\omega$ belongs to the
spectrum of our model then $h(\omega)=0$;
$h(\omega)$ does not have any singularities.
It is tempting to wright that
$\rho (\omega) = (1/\pi)(d/d\omega) {\rm Im}(\ln h(\omega))$. However, this
is incorrect. There is one thing we have to take care of: for a
given eigenvalue $\omega\ne 0$ the function $h$ has two zeros (at $\omega$ 
and $-\omega$), for $\omega=0$ it has a second order zero.
The following is the correct expression for the density of states:
\begin{equation}
\rho={1\over\pi}{d\over d\omega}{\rm Im}\left({1\over 2}\ln h(\omega)\right).
\end{equation}
The factor of $(1/2)$ removes the impact of extra zeros of $h(\omega)$.

Then:
\begin{equation}
\rho={1\over 2\pi}{d\over d\omega}{\rm Im}\left(\ln\omega+\right.
\end{equation}
$$
\ln\left[\omega - 2\Delta^2\sum_{k}{1\over\omega+\epsilon_k}
+{1\over\omega-\epsilon_k}\right]+
$$
$$
\left.\sum_{k}\ln(\omega+\epsilon_k) 
+\sum_{k}\ln(\omega-\epsilon_k)\right).
$$
Expression in square brackets is evaluated in the following way:
\begin{equation}
\omega-2\Delta^2\sum_{k}{1\over\omega+\epsilon_k}+{1\over\omega-\epsilon_k}
=
\end{equation}
$$
\omega-\Delta^2{2L\over\pi v_F}\int_{-\Omega}^{\Omega}
{d\epsilon\over\omega + \epsilon}+{d\epsilon\over\omega-\epsilon} =
\omega-i\Gamma, 
$$
where bandwidth $\Omega$ and coupling $\Gamma$ are defined as follows:
\begin{equation}
\Omega={\pi v_F\over a},
\end{equation}
\begin{equation}
\Gamma={J^{2}_{x}\over\pi a v_F}\ll\Omega.
\end{equation}

After straightforward calculations the density of states is found to be 
\begin{equation}
\rho(\omega)=\rho_0 + {1\over 2\pi}{\Gamma\over \omega^2+\Gamma^2} +
{1\over2}\delta(\omega),
\end{equation}
where $\rho_0$ is the density of states without the impurity.

Thus, the impurity free energy is:
\begin{equation}
{\cal F}=-T\int_{-\Omega}^{\Omega}{1\over 2\pi}{\Gamma\over \omega^2+\Gamma^2}
\ln\left(1+e^{-{\omega/ T}}\right) d\omega - {T\over 2}\ln 2.
\label{free}
\end{equation}
This free energy coincides with the free energy 
found in I up to constant. From (\ref{free}) zero temperature value 
of the impurity entropy is found to be $(1/2)\ln 2$. But, as mentioned
above, (\ref{free}) is not valid below $T_L$.

\section{Conclusions}

Fractional entropy paradox does not exist for the finite
system. The presented calculations provide us with the direct check of
that.
In the case of finite system Kondo impurity can be described in
terms of free fermions. It was proven above that none of them is of Majorana
type. Using this representation we found the entropy $S$ for arbitrary $T$
and $L$. As expected, $S$ shows a cross-over at $T_L=v_F/L$.

We derived high temperature expression for the impurity free energy and
found that it coincides with the free energy obtained in $L=\infty$ limit.

\section{Acknowledgments}
Author expresses his sincere gratitude to F. Guinea without whom this work
could not be possible and to D. Arovas whose encouragement and advice was
extremely helpful. Author would like to thank D. Cox and I. Affleck who
draw his attention to the important results presented in \cite{affleck}.


\begin{thebibliography}{99}
\bibitem{expl} The first limit corresponds to the cooling experiment
with a sample of fixed size. By cooling it low enough we have to see log of
some integer. The second limit corresponds to another set of experiments
when many samples of increasing size are studied at the fixed temperature.
In this case there is no fundamental restriction on the limiting value of
the entropy as well as that limiting value does not have any fundamental
meaning.
\end{thebibliography}
\end{document}